\documentclass[preprint,showpacs,preprintnumbers,amsmath,amssymb]{revtex4}

\usepackage{graphicx,amsfonts}
\usepackage{epsfig}
\usepackage{dcolumn}
\usepackage{bm}
\begin{document}

\def\nonu{\nonumber}
\def\rf#1{(\ref{eq:#1})}
\def\lab#1{\label{eq:#1}} 
\def\br{\begin{eqnarray}}
\def\er{\end{eqnarray}}
\def\be{\begin{equation}}
\def\ee{\end{equation}}
\def\0{\nonumber}
\def\lb{\lbrack}
\def\rb{\rbrack}
\def\({\left(}
\def\){\right)}
\def\v{\vert}
\def\bv{\bigm\vert}
\def\lskip{\vskip\baselineskip\vskip-\parskip\noindent}
\relax
\newcommand{\nit}{\noindent}
\newcommand{\ct}[1]{\cite{#1}}
\newcommand{\bi}[1]{\bibitem{#1}}
\def\a{\alpha}
\def\b{\beta}
\def\ca{{\cal A}}
\def\cm{{\cal M}}
\def\cn{{\cal N}}
\def\cf{{\cal F}}
\def\d{\delta} 
\def\D{\Delta}
\def\eps{\epsilon}
\def\g{\gamma}
\def\G{\Gamma}
\def\grad{\nabla}
\def\h{ {1\over 2}  }
\def\hc{\hat{c}}
\def\hd{\hat{d}}
\def\hg{\hat{g}}
\def\hp{ {+{1\over 2}}  }
\def\hm{ {-{1\over 2}}  }
\def\k{\kappa}
\def\l{\lambda}
\def\L{\Lambda}
\def\lg{\langle}
\def\m{\mu}
\def\n{\nu}
\def\o{\over}
\def\om{\omega}
\def\O{\Omega}
\def\p{\phi}
\def\pa{\partial}
\def\pr{\prime}
\def\ra{\rightarrow}
\def\rh{\rho}
\def\rg{\rangle}
\def\s{\sigma}
\def\t{\tau}
\def\th{\theta}
\def\ti{\tilde}
\def\wti{\widetilde}
\def\inte{\int dx }
\def\xb{\bar{x}}
\def\yb{\bar{y}}

\def\tr{\mathop{\rm tr}}
\def\Tr{\mathop{\rm Tr}}
\def\partder#1#2{{\partial #1\over\partial #2}}
\def\ds{{\cal D}_s}
\def\wtwo{{\wti W}_2}
\def\lie{{\cal G}}
\def\alie{{\widehat \lie}}
\def\dlie{{\cal G}^{\ast}}
\def\elie{{\widetilde \lie}}
\def\edlie{{\elie}^{\ast}}
\def\hlie{{\cal H}}
\def\wlie{{\widetilde \lie}}

\def\rlx{\relax\leavevmode}
\def\inbar{\vrule height1.5ex width.4pt depth0pt}
\def\IZ{\rlx\hbox{\sf Z\kern-.4em Z}}
\def\IR{\rlx\hbox{\rm I\kern-.18em R}}
\def\IC{\rlx\hbox{\,$\inbar\kern-.3em{\rm C}$}}
\def\one{\hbox{{1}\kern-.25em\hbox{l}}}

\def\PRL#1#2#3{{\sl Phys. Rev. Lett.} {\bf#1} (#2) #3}
\def\NPB#1#2#3{{\sl Nucl. Phys.} {\bf B#1} (#2) #3}
\def\NPBFS#1#2#3#4{{\sl Nucl. Phys.} {\bf B#2} [FS#1] (#3) #4}
\def\CMP#1#2#3{{\sl Commun. Math. Phys.} {\bf #1} (#2) #3}
\def\PRD#1#2#3{{\sl Phys. Rev.} {\bf D#1} (#2) #3}
\def\PLA#1#2#3{{\sl Phys. Lett.} {\bf #1A} (#2) #3}
\def\PLB#1#2#3{{\sl Phys. Lett.} {\bf #1B} (#2) #3}
\def\JMP#1#2#3{{\sl J. Math. Phys.} {\bf #1} (#2) #3}
\def\PTP#1#2#3{{\sl Prog. Theor. Phys.} {\bf #1} (#2) #3}
\def\SPTP#1#2#3{{\sl Suppl. Prog. Theor. Phys.} {\bf #1} (#2) #3}
\def\AoP#1#2#3{{\sl Ann. of Phys.} {\bf #1} (#2) #3}
\def\PNAS#1#2#3{{\sl Proc. Natl. Acad. Sci. USA} {\bf #1} (#2) #3}
\def\RMP#1#2#3{{\sl Rev. Mod. Phys.} {\bf #1} (#2) #3}
\def\PR#1#2#3{{\sl Phys. Reports} {\bf #1} (#2) #3}
\def\AoM#1#2#3{{\sl Ann. of Math.} {\bf #1} (#2) #3}
\def\UMN#1#2#3{{\sl Usp. Mat. Nauk} {\bf #1} (#2) #3}
\def\FAP#1#2#3{{\sl Funkt. Anal. Prilozheniya} {\bf #1} (#2) #3}
\def\FAaIA#1#2#3{{\sl Functional Analysis and Its Application} {\bf #1} (#2)
#3}
\def\BAMS#1#2#3{{\sl Bull. Am. Math. Soc.} {\bf #1} (#2) #3}
\def\TAMS#1#2#3{{\sl Trans. Am. Math. Soc.} {\bf #1} (#2) #3}
\def\InvM#1#2#3{{\sl Invent. Math.} {\bf #1} (#2) #3}
\def\LMP#1#2#3{{\sl Letters in Math. Phys.} {\bf #1} (#2) #3}
\def\IJMPA#1#2#3{{\sl Int. J. Mod. Phys.} {\bf A#1} (#2) #3}
\def\AdM#1#2#3{{\sl Advances in Math.} {\bf #1} (#2) #3}
\def\RMaP#1#2#3{{\sl Reports on Math. Phys.} {\bf #1} (#2) #3}
\def\IJM#1#2#3{{\sl Ill. J. Math.} {\bf #1} (#2) #3}
\def\APP#1#2#3{{\sl Acta Phys. Polon.} {\bf #1} (#2) #3}
\def\TMP#1#2#3{{\sl Theor. Mat. Phys.} {\bf #1} (#2) #3}
\def\JPA#1#2#3{{\sl J. Physics} {\bf A#1} (#2) #3}
\def\JSM#1#2#3{{\sl J. Soviet Math.} {\bf #1} (#2) #3}
\def\MPLA#1#2#3{{\sl Mod. Phys. Lett.} {\bf A#1} (#2) #3}
\def\JETP#1#2#3{{\sl Sov. Phys. JETP} {\bf #1} (#2) #3}
\def\JETPL#1#2#3{{\sl  Sov. Phys. JETP Lett.} {\bf #1} (#2) #3}
\def\PHSA#1#2#3{{\sl Physica} {\bf A#1} (#2) #3}
\def\PHSD#1#2#3{{\sl Physica} {\bf D#1} (#2) #3}

\title{Classical Integrable $N=1$ and $N= 2$ Super Sinh-Gordon Models with Jump Defects }

\author{J.F. Gomes,  L.H. Ymai and A.H. Zimerman}%

\affiliation{
Instituto de F\'\i sica Te\'orica,
Universidade Estadual Paulista\\ Rua Pamplona, 145
01405-900 - S\~ao Paulo, SP, Brazil
}

\date{\today}

\begin{abstract}
The structure of integrable field theories in the presence  of jump defects is discussed in 
 terms of boundary functions under the Lagrangian formalism.  
 Explicit examples of  bosonic and fermionic theories are considered.  In particular, 
 the boundary functions for the $N=1$ and $N=2$ super sinh-Gordon models are constructed and shown to 
  generate the Backlund transformations for its soliton solutions.  As a new and interesting example,  a solution with an incoming boson and an outgoing fermion  for the $N=1$ case is presented.  The resulting  integrable models are shown to be
invariant under supersymmetric transformation.

\end{abstract}

\maketitle
\section{Introduction}
The classical Lagrangian formulation of a class of relativistic integrable field theories 
admiting certain discontinuities (defects) has been 
studied recently  \cite{bowcock1} -  \cite{corrigan}.
In particular, in ref. \cite{bowcock1} the authors  have considered  a field theory in which different soliton 
solutions of the sine-Gordon model  are linked in such a way that the integrability is preserved.
The integrability of the total system imposes severe constraints specifying  the possible
types of defects.  These are characterized by Backlund transformations which are known to  
connect two different soliton solutions. 

The presence of the defect indicates the breakdown of space isotropy and henceforth of momentum  conservation.
The key ingredient to classify integrable defects is to impose certain  first order  differential 
relations between the different solutions (Backlund transformation).  
This introduces certain boundary functions (BF) which are 
specific of each integrable model  and leads to the  conservation of the total momentum.

Here, we analize the structure of the possible boundary terms for various cases.
 We first  consider, for pedagogical purposes, the pure bosonic case  studied in refs.  \cite{bowcock1} -  \cite{corrigan}  and derive, according to them,  the border functions 
 by imposing  conservation of the total momentum.  Next, we consider  a pure fermionic theory  and propose  Backlund transformation  in terms
 of an auxiliary  fermionic  non local field.    Such  structure is then generalized to include  both bosonic and fermionic fields.  In
 particular we construct boundary functions  for the  $N=1$  supersymmetric sinh-Gordon model  and show that it leads to the  Backlund
 transformation proposed  in \cite{chai}.  
 
 We extend the formalism to the $N=2$ supersymmetric sinh-Gordon model \cite{inprep}. This system was originally proposed in \cite{ik} and \cite{ku} (see also \cite{nepo}).  Later in \cite{nosso1} and \cite{nosso2} a systematic algebraic aproach was developed.
 
\section{General Formalism - Bosonic Case}
In this section we introduce the Lagrangian approach  proposed in \cite{bowcock1}. 
Consider a system  described  by 
\begin{eqnarray} 
{\cal{L}}=\theta(-x){\cal{L}}_{p=1}+\theta(x){\cal{L}}_{p=2}+\delta(x){\cal{L}}_{D},
\label{1.1}
\end{eqnarray}
where ${\cal L}_p(\phi_p, \pa_{\mu} \phi_p) = {1\o 2} (\pa_x \phi_p)^2 -  {1\o 2} (\pa_t \phi_p)^2 - V(\phi_p)$ 
describes  a set of fields denoted by $\phi_1$ for $x<0$ and 
$\phi_2$ for $x>0$.  A defect  placed at $x=0$,  is described by 
\begin{eqnarray} 
{\cal L}_D = {1\o 2} ( \phi_2 \pa_t \phi_1 - \phi_1 \pa_t \phi_2) + B_0
\label{1.2}
\end{eqnarray}
where $B_0$ is the border function.  The equations of motion are therefore given by
\begin{eqnarray}
\partial_{x}^{2}\phi_{1}-\partial_{t}^{2}\phi_{1} &=& \pa_{\phi_1}V(\phi_1), \quad \quad x<0
\nonumber\\
\partial_{x}^{2}\phi_{2}-\partial_{t}^{2}\phi_{2} &=& \pa_{\phi_2}V(\phi_2), \quad \quad x>0
\nonumber\\ 
\label{1.3}
\end{eqnarray}
and  $x=0$,
\begin{eqnarray}
\pa_x \phi_1 - \pa_t \phi_2 &=& - \pa_{\phi_1} B_0, \nonumber \\
\pa_x \phi_2 - \pa_t \phi_1 &=&  \pa_{\phi_2} B_0, \nonumber \\
\label{1.4}
\end{eqnarray}
The momentum is 
\begin{eqnarray}
P = \int_{-\infty}^{0} \pa_x \phi_1 \pa_t \phi_1 + \int^{\infty}_{0} \pa_x \phi_2 \pa_t \phi_2.
\label{1.5}
\end{eqnarray}
Acting with time derivative and inserting eqns. of motion (\ref{1.3}) we find
\begin{eqnarray}
{{dP}\o {dt}} &=& \int_{-\infty}^{0} ({1\o 2} \pa_x (\pa_t \phi_1)^2 + {1\o 2} \pa_x (\pa_x\phi_1)^2 - \pa_x \phi_1 {{\d V_1}\o {\d \phi_1}}
)dx \nonumber \\
&+& \int^{\infty}_{0} ({1\o 2} \pa_x (\pa_t \phi_2)^2 + {1\o 2} \pa_x (\pa_x\phi_2)^2 - \pa_x \phi_2{{\d V_2}\o {\d \phi_2}})dx 
\label{1.6}
\end{eqnarray}
Using eqns. (\ref{1.4}) after integration, we find
\begin{eqnarray}
{{dP}\o {dt}} &=& [-{{\pa B_0}\o {\pa \phi_+}} \dot {\phi_+} + {{\pa B_0}\o {\pa \phi_-}} \dot {\phi_-} + 
{1\o 2} ( {{\pa B_0}\o {\pa \phi_1}})^2 - {1\o 2} ( {{\pa B_0}\o {\pa \phi_2}})^2 -V_1 +V_2 ]|_{x=0}
\label{1.7}
\end{eqnarray}
where $\phi_{\pm} = \phi_1 \pm \phi_2$ and $\dot {\phi_{\pm}}= \pa_t \phi_{\pm}$.  
If the border function factorizes into $B_0 = B_0^{+} (\phi_+) + B_0^{-} (\phi_-)$, 
the modified momentum ${\cal P} = P + \(B_0^{+}  - B_0^{-}\)|_{x=0} $ is  conserved  provided   $B_0$  satisfies its defining condition, i.e.,
\begin{eqnarray}
[{1\o 2} ( {{\pa B_0}\o {\pa \phi_1}})^2 - {1\o 2} ( {{\pa B_0}\o {\pa \phi_2}})^2 -V_1 +V_2 ]|_{x=0} = 0
\label{1.8}
\end{eqnarray}
 
Let us illustrate  the  above structure by first considering the free massive  bosonic theory for which $V_p = {1\o 2} m^2 \phi_p^2$.   
\begin{eqnarray}
{1\o 2} ( {{\pa B_0}\o {\pa \phi_1}})^2 - {1\o 2} ( {{\pa B_0}\o {\pa \phi_2}})^2 = {{m^2}\o 2} (\phi_1^2 - \phi_2^2) 
= {{m^2}\o 2}\phi_{+}  \phi_{-}
\label{1.9}
\end{eqnarray}
The solution is  easely found to be
\begin{eqnarray}
B_0 = - {{m\b^2}\o{4}} \phi_-^2 - {{m}\o{4\b^2} }\phi_+^2
\label{1.10}
\end{eqnarray}
and $\b^2$ denotes a free  (spectral) parameter.

As second  example, consider the sinh-Gordon model for which $V_p = 4 m^2 \cosh (2\phi_p)$. The
defining eqn. (\ref{1.8})  indicates the natural decomposition
\begin{eqnarray}
{1\o 2} ( {{\pa B_0}\o {\pa \phi_1}})^2 - {1\o 2} ( {{\pa B_0}\o {\pa \phi_2}})^2 
&=& 4m^2 (\cosh(2\phi_1)  - \cosh(2\phi_2) ) \nonumber \\
&=& 8m^2 \sinh(\phi_+) \sinh (\phi_-)
\label{1.11}
\end{eqnarray}
yielding 
\begin{eqnarray}
B_0 = -m\b^2 \cosh (\phi_-) - {{4m}\o {\b^2}}\cosh(\phi_+)
\label{1.12}
\end{eqnarray}
and hence we rederive from (\ref{1.4}) the Backlund transformation for the sinh-Gordon model
\begin{eqnarray}
\pa_x \phi_1 - \pa_t \phi_2 &=& m\b^2 \sinh(\phi_-) + {{4m}\o {\b^2}} \sinh(\phi_+), \nonumber \\
\pa_x \phi_2 - \pa_t \phi_1 &=& m\b^2 \sinh(\phi_-) - {{4m}\o {\b^2}} \sinh(\phi_+) , 
\label{1.13}
\end{eqnarray}

\section{Fermions and the $N=1$ Super sinh-Gordon Model}

Before  discussing the Super sinh-Gordon Model let us consider the pure fermionic prototype described by the Lagrangian density
\begin{eqnarray} 
{\cal{L}}_p= \bar \psi_p \pa_t \bar \psi_p - \bar \psi_p \pa_x \bar \psi_p +  \psi_p \pa_t \psi_p +  \psi_p \pa_x  \psi_p + W(\psi_p, \bar \psi_p)
\label{2.1}
\end{eqnarray}
where, for the  free fermionic theory, $W(\psi_p, \bar \psi_p)= 2m \bar \psi_p \psi_p$.
For the half line, $x<0$ or $x>0$   the equations of motion are given by
\begin{eqnarray} 
\pa_x \psi_p + \pa_t \psi_p = -{1\o 2} \pa_{\psi_p} W_p, \quad \quad \pa_x \bar \psi_p - \pa_t \bar \psi_p = {1\o 2} \pa_{\bar \psi_p} W_p
\label{2.2}
\end{eqnarray}
according to $p=1$ or $2$ respectively.
 
Let us   propose  the following Backlund transformation  
\begin{eqnarray} 
\psi_1 + \psi_2 =  -i \b \sqrt{m} f_1 = \pa_{\psi_1} B_1, \quad 
\quad \bar \psi_1 - \bar \psi_2 = -{{2i \sqrt{m}}\o {\b}} f_1 = -\pa_{\bar \psi_1} B_1, \quad 
\label{2.3}
\end{eqnarray}
where $f_1$ satisfies
\begin{eqnarray} 
 \dot {f_1} &=& -{{i\b \sqrt{m}}\o 2} (\psi_1 - \psi_2) + {{i}\o {2\b}}\sqrt{m} (\bar \psi_1 +\bar \psi_2) = -{1\o 4}\pa_{f_1} B_1, \nonumber \\
 \pa_x {f_1} &=& {{i\b \sqrt{m}}\o 2} (\psi_1 - \psi_2) + {{i}\o {2\b}}\sqrt{m} (\bar \psi_1 +\bar \psi_2)
\label{2.3a}
\end{eqnarray}
written in terms of  a border function $B_1= B_1 (\bar \psi_1, \bar \psi_2,\psi_1,\psi_2, f_1)$, 
which now, due to the Grassmanian character of the fermions,
depends upon  the non local fermionic field $f_1$.
 By considering the Lagrangian system (\ref{1.1}) with ${\cal {L}}_p$ given by (\ref{2.1}) and 
\begin{eqnarray} 
{\cal L}_D = -\psi_1 \psi_2 - \bar \psi_1 \bar \psi_2 +2 f_1 \pa_t f_1 + B_1 (\bar \psi_1, \bar \psi_2,\psi_1,\psi_2, f_1)
\label{2.2a}
\end{eqnarray} 
 construct the momentum to be
\begin{eqnarray}
{P} &=& \int_{-\infty}^{0} ( - \bar \psi_1 \pa_x \bar \psi_1 - \psi_1 \pa_x  \psi_1)dx 
+ \int^{\infty}_{0} (- \bar \psi_2 \pa_x \bar \psi_2 - \psi_2 \pa_x  \psi_2)dx 
\label{2.3ab}
\end{eqnarray} 
 In  analising its   conservation, we find 
 \begin{eqnarray}
{{dP}\o {dt}} &=& [-W_1 +W_2  - \bar \psi_1 \pa_t \bar \psi_1 + \bar \psi_2 \pa_t \bar \psi_2 
-  \psi_1 \pa_t  \psi_1 +  \psi_2 \pa_t  \psi_2 ]|_{x=0}
\label{2.4}
\end{eqnarray}
 after using equations  of motion  (\ref{2.2})  to 
 eliminate time derivatives  and integrating  over $x$.  Using the Backlund transformation (\ref{2.3}) and (\ref{2.3a}),
 eqn. (\ref{2.4}) becomes
 \begin{eqnarray}
{{dP}\o {dt}} &=& -[W_1 -W_2  - \pa_{\bar \psi_1} B_1 \dot {\bar \psi_1} +  \pa_{ \psi_1} B_1 \dot { \psi_1} \nonumber \\
&-&\pa_{\bar \psi_2} B_1 \dot {\bar \psi_2} +  \pa_{ \psi_2} B_1 \dot { \psi_2} 
+ \pa_t(\bar \psi_2 \bar \psi_1) - \pa_t( \psi_2 \psi_1) ]|_{x=0}
\label{2.5}
\end{eqnarray}
If we assume   the border function to  decompose as $B_1 = B_1^+ (\bar \psi_+, f_1) + B_1^- ( \psi_-, f_1)$, where $\bar \psi_{\pm} = \bar \psi_1 \pm \bar \psi_2, \quad \psi_{\pm} =  \psi_1 \pm  \psi_2$,
 the   modified momentum  
 \begin{eqnarray}
 {\cal P} = P  +[ \bar \psi_2 \bar \psi_1- \psi_2 \psi_1 + B_1^+ - B_1^-]_{x=0}
 \label{2.6}
\end{eqnarray} 
is conserved provided  
\begin{eqnarray}
[W_2 -W_1 - \pa_{f_1} B_1^- \dot{f_1} + \pa_{f_1} B_1^+ \dot{f_1}]|_{x=0}  = 0
\label{2.7}
\end{eqnarray} 
For the free fermi fields system (\ref{2.1})-(\ref{2.2}) eqn. (\ref{2.7}) becomes
\begin{eqnarray}
{1\o 2} (\pa_{f_1} B_1^+)   (\pa_{f_1} B_1^-) = 2m (\bar \psi_1 \psi_1 - \bar \psi_2 \psi_2)
\label{2.8}
\end{eqnarray}
The solution is 
\begin{eqnarray}
B_1 = -{{2i}\o {\b}}\sqrt{m}f_1 \bar \psi_+ + i\b \sqrt{m} f_1\psi_-
\label{2.9}
\end{eqnarray}

Let us now consider the super sinh-Gordon model described by
\begin{eqnarray} 
{\cal{L}}_p&=& {1\o 2} (\pa_x \phi_p)^2 -{1\o 2} (\pa_t \phi_p)^2 
+ \bar \psi_p \pa_t \bar \psi_p - \bar \psi_p \pa_x \bar \psi_p +  \psi_p \pa_t \psi_p \nonumber \\
&+&  
\psi_p \pa_x  \psi_p + V( \phi_p) +W(\phi_p, \psi_p, \bar \psi_p)
\label{2.10}
\end{eqnarray}
and 
\begin{eqnarray} 
{\cal{L}}_D&=& {1\o 2} ( \phi_2 \pa_t \phi_1 - \phi_1 \pa_t \phi_2) -\psi_1 \psi_2 - 
\bar \psi_1 \bar \psi_2  + 2f_1 \pa_t f_1 \nonumber \\
&+&B_0(\phi_1,\phi_2) + B_1(\phi_1, \phi_2,\bar \psi_1, \bar \psi_2,\psi_1,\psi_2, f_1)
\label{2.11}
\end{eqnarray}
where $V_p = 4m^2 \cosh(2\phi_p) $ and $ W_p = 8m \bar \psi_p\psi_p \cosh (\phi_p)$.

Propose the following Backlund transformation  \cite{ymai},
\begin{eqnarray}
\pa_x \phi_1 - \pa_t \phi_2 = -\pa_{\phi_1} (B_0 +B_1), 
\quad \quad \pa_x \phi_2 - \pa_t \phi_1 = \pa_{\phi_2} (B_0 +B_1)\nonumber \\
\psi_1 + \psi_2 =   \pa_{\psi_1} B_1, 
\quad \bar \psi_1 - \bar \psi_2 =  -\pa_{\bar \psi_2} B_1, \quad \dot f_1 = -{1\o 4}\pa_{f_1} B_1
\label{2.12}
\end{eqnarray}
Assuming the decomposition
\begin{eqnarray}
B_0 = B_0^+(\phi_+) + B_0^-(\phi_-), \quad \quad 
B_1 = B_1^+ (\phi_+, \bar \psi_+, f_1) + B_1^- ( \phi_-, \psi_-, f_1)
\label{2.13}
\end{eqnarray}
we find that the modified momentum 
\begin{eqnarray}
{\cal P} = P + [B_0^+(\phi_+) - B_0^-(\phi_-) +  B_1^+ (\phi_+,\bar \psi_+, f_1) - B_1^- ( \phi_-, \psi_-, f_1)- \bar \psi_1 \bar \psi_2 + \psi_1 \psi_2]_{x=0}
\nonu \\
\label{2.14}
\end{eqnarray}
is conserved provided the border functions $B_0$ and $B_1$ satisfy
\begin{eqnarray}
 W_1 - W_2 &=& {1\o 2} (\pa_{f_1} B_1^+) (\pa_{f_1} B_1^-) 
 + 2 (\pa_{\phi_+} B_0^+) (\pa_{\phi_-} B_1^-) + 2  (\pa_{\phi_-} B_0^-)(\pa_{\phi_+} B_1^+) 
\nonumber \\
V_1 - V_2 &=& {1\o 2} ( {{\pa B_0}\o {\pa \phi_1}})^2 - {1\o 2} ( {{\pa B_0}\o {\pa \phi_2}})^2 
\label{2.15}
\end{eqnarray}
The solution for (\ref{2.15}) is found to be
\begin{eqnarray}
B_0 &=& -m \b^2 \cosh( \phi_-) -{{4m}\o {\b^2}} \cosh (\phi_+), \nonumber \\
B_1 &=& -{{4i}\o {\b}} \sqrt{m} \cosh({{\phi_+}\o {2}}) f_1 \bar \psi_+ + 2i\b \sqrt{m} \cosh({{\phi_-}\o {2}}) f_1  \psi_-
\label{2.16}
\end{eqnarray}
The boundary functions $B_0$ and $B_1$ in (\ref{2.16}) generate the Backlund  transformation  which 
 agrees with the  one proposed in \cite{chai}  for the $N=1$ super sinh-Gordon model.  
 The equations of motion obtained from (\ref{2.10}) and (\ref{2.11}) are verified to be invariant under the supersymmetry transformation
 \br
 \d \bar \psi_p = \eps \pa_z \phi_p, \quad \d \phi_p = \eps \bar \psi_p, \quad \d \psi_p = 2 \eps m \sinh \phi_p
 \er
 where $\pa_z = 1/2 (\pa_x + \pa_t)$.
 
 In ref. \cite{pla}  the general  soliton solutions for the $N=1$ super sinh-Gordon model 
  were  constructed using vertex functions techniques and in \cite{ymai} 
 different solutions  were analysed in the context of the Backlund  transformation.
Besides those examples discussed in \cite{ymai}, a new and interesting case containing {\it an incoming boson and an outcoming  fermion} is given when $\phi_1 \neq 0, \psi_1 = \bar \psi_1 =0$ and $ \phi_2 = 0, \psi_2, \bar \psi_2 \neq 0$.  Under such conditions the Backlund transformation (\ref{2.12}) reads,
\br
\pa_x \phi_1 &=& -2 (\s +{{1}\o {\s}})\sinh  \phi_1, \quad \quad  \pa_t \phi_1 = -2 (\s -{{1}\o {\s}})\sinh \phi_1,\nonu \\
\psi_2 &=& 2\sqrt{{{2}\o {\s}}} \cosh \( {{\phi_1}\o 2}\) f_1, \quad \quad \bar \psi_2 = 2\sqrt{{{2} {\s}}} \cosh \({{\phi_1}\o 2}\) f_1, \nonu \\
\pa_t f_1 &=& 2(\s - {{1}\o {\s}})\cosh^2 \( {{\phi_1}\o 2}\) f_1, \quad \quad \pa_x f_1 = 2(\s + {{1}\o {\s}})\cosh^2 \({{\phi_1}\o 2}\) f_1
\label{back}
\er
where we took $m=1$ and $\s = -{{2}\o {\b^2}}$.  As solution of eqns. (\ref{back}) we find
\br
\phi_1 = ln \({{1+{{b_1}\o {2}}\rho}\o {1- {{b_1}\o{2}}\rho}}\), \quad \quad f_1 = \eps \rho^{-1} \sqrt{1- {{b_1^2}\o 4}\rho^2}
\er
where $\rho = \exp \(-2\s (x+t) -{{2}\o {\s}}(x-t)\)$, $\eps$ is a grassmaniann constant and $\psi_2$ and $\bar \psi_2$ are obtained from the second eqn. (\ref{back}) after  inserting the above $\phi_1$ and $f_1$.

The integrability of the model is verified by construction of  the Lax pair representation of the 
equations of motion.  This is achieved by
splitting the space into two overlapping regions, namely, $x\leq b$ and $x \geq a$ with $a<b$ 
and defining  a corresponding Lax pair  within each of them.   The integrability is ensured 
 by the existence of a gauge transformation 
relating the  two sets of Lax pairs within the overlapping region.  
In ref. \cite{ymai} we have  explicitly constructed such gauge transformation for the $N=1$
super sinh-Gordon model in terms of the $SL(2,1)$
affine Lie algebra.

\section{$N=2$ Super sinh-Gordon Model}

The Lagrangian density  for the $N=2$ Super sinh-Gordon Model is given by ( see for instance \cite{nepo})
\br
{\cal L}_p &=& {1\o 2} (\pa_x \phi_p)^2 -{1\o 2} (\pa_t \phi_p)^2 + 2\bar \psi_p \pa_t \bar \psi_p + 2\bar \psi_p \pa_x \bar \psi_p  + 2\psi_p \pa_t  \psi_p - 2\psi_p \pa_x  \psi_p \nonu \\
&-& {1\o 2} (\pa_x \varphi_p)^2 +{1\o 2} (\pa_t \varphi_p)^2 -2 \bar
\chi_p \pa_t \bar \chi_p - 2\bar \chi_p \pa_x \bar \chi_p
-2 \chi_p \pa_t  \chi_p + 2\chi_p \pa_x  \chi_p \nonu \\
&-& 16 (\psi_p \bar \psi_p + \chi_p \bar \chi_p) \cosh \varphi_p \cosh \phi_p - 4 \cosh (2 \varphi_p) \nonu \\
&+& 16 (\psi_p \bar \chi_p + \chi_p \bar \psi_p) \sinh \varphi_p \sinh \phi_p + 4 \cosh (2\phi_p)
\label{n2}
\er
whose equations of motion are  invariant under the supersymmetry transformations
\br
\d (\phi_p \pm \varphi_p)= 2 (\psi_p \mp \chi_p) \eps_{\pm},  \quad \d (\psi_p \pm \chi_p) = - \pa_z (\phi_p \mp \varphi_p)\eps_{\pm}
\label{s}
\er
and 
\br
\d (\bar \psi_p \pm \bar \chi_p) = 2 \sinh (\phi_p \pm \varphi_p) \eps_{\mp}
 \label{s1}
 \er

Inspired from the $N=1$ case (see eqn. (\ref{2.11})) we propose the following  Lagrangian description for the defect
\br {\cal L}_{D} &=& {{1\o 2}}(\phi_2 \pa_t \phi_1 - \phi_1 \pa_t
\phi_2) -2\psi_1 \psi_2 - 2\bar \psi_1 \bar \psi_2
+  f_1 \pa_t f_2 \nonu \\ 
&-& {1\o 2} (\varphi_2 \pa_t \varphi_1 - \varphi_1 \pa_t \varphi_2)
+2\chi_1 \chi_2 + 2\bar \chi_1 \bar \chi_2
+  f_2 \pa_t f_1 \nonu \\ 
&+& B_0(\phi_p, \varphi_p) + B_1 (\phi_p, \varphi_p, \psi_p, \chi_p, \bar \psi_p, \bar \chi_p, f_1, f_2)
\label{def}
\er
leading to the Backlund transformation
\br
\pa_x \phi_1 - \pa_t \phi_2 &=&  - \pa_{\phi_1} B,\nonu \\
\pa_x \phi_2 - \pa_t \phi_1 &=&   \pa_{\phi_2} B, \nonu \\
\pa_x \varphi_1 - \pa_t \varphi_2 &=&   \pa_{\varphi_1} B, \nonu \\
\pa_x \varphi_2 - \pa_t \varphi_1 &=&   -\pa_{\varphi_2} B, \nonu \\
\psi_1 -\psi_2 &=& -{1\o 2}\pa_{\psi_1} B=-{1\o 2}\pa_{\psi_2} B\nonu \\
\chi_1 -\chi_2 &=& {1\o 2}\pa_{\chi_1} B={1\o 2}\pa_{\chi_2} B\nonu \\
\bar \psi_1 + \bar \psi_2 &=& {1\o 2}\pa_{\bar \psi_1}B=-{1\o 2}\pa_{\bar \psi_2}B, \nonu \\
\bar \chi_1 + \bar \chi_2 &=& -{1\o 2}\pa_{\bar \chi_1}B={1\o 2}\pa_{\bar \chi_2}B, \nonu \\
\dot f_1 &=& -{1\o 2} \pa_{f_2} B, \nonu \\
\dot f_2 &=& -{1\o 2} \pa_{f_1} B
\label{d}
\er
where $B = B_0 + B_1$.  The canonical momentum 
\br
P &=& \int_{-\infty}^0 \( \pa_x \phi_1 \pa_t \phi_1 -2 \bar \psi_1 \pa_x \bar \psi_1 -2 \psi_1 \pa_x \psi_1 \right. \nonu \\
 &-& \left. \pa_x \varphi_1 \pa_t \varphi_1 + 2 \bar \chi_1 \pa_x \bar \chi_1 + 2 \chi_1 \pa_x \chi_1\) dx \nonu \\
&+&
 \int^{\infty}_0 \( \pa_x \phi_2 \pa_t \phi_2 -2 \bar \psi_2 \pa_x \bar \psi_2 -2 \psi_2 \pa_x \psi_2 \right. \nonu \\
 &-&\left. \pa_x \varphi_2 \pa_t \varphi_2 +2 \bar \chi_2 \pa_x \bar \chi_2 + 2\chi_2 \pa_x \chi_2\)dx 
 \label{mom}
 \er
is no longer conserved.  Proposing the modified momentum to be 
\br
{\cal P} &=& P + [B_0^{(+)} - B_0^{(-)}+ B_1^{(+)} - B_1^{(-)} 
+ 2\bar \psi_1 \bar \psi_2 -2\psi_1 \psi_2 - 2\bar \chi_1 \bar
\chi_2 + 2\chi_1 \chi_2]_{x=0} \nonu 
\er
which is conserved provided the border function satisfies
\br
   \pa_{\phi_+} B_0^{(+)} \pa_{\phi_-}B_0^{(-)} -  \pa_{\varphi_+} B_0^{(+)} \pa_{\varphi_-}B_0^{(-)}  
  = 4 \sinh \phi_+ \sinh \phi_- - 4 \sinh \varphi_+ \sinh \varphi_-, \nonu \\
\label{b0}
\er
and 
\br
& & \pa_{\phi_+} B_0^{(+)} \pa_{\phi_-} B_1^{(-)} + \pa_{\phi_-} B_0^{(-)} \pa_{\phi_+} B_1^{(+)}  
- \pa_{\varphi_+} B_0^{(+)} \pa_{\varphi_-} B_1^{(-)} - \pa_{\varphi_-} B_0^{(-)} \pa_{\varphi_+} B_1^{(+)} \nonu \\
&+&\pa_{\phi_+} B_1^{(+)} \pa_{\phi_-} B_1^{(-)} - \pa_{\varphi_+} B_1^{(+)} \pa_{\varphi_-} B_1^{(-)} 
- {1\o 2} (\pa_{f_1} B_1^{(-)} \pa_{f_2}B_1^{(+)} + \pa_{f_2} B_1^{(-)} \pa_{f_1}B_1^{(+)}) \nonu \\
&=&
-2(\psi_+ \bar \psi_+ + \psi_- \bar \psi_- +\chi_+ \bar \chi_+ + \chi_- \bar \chi_-)\L_- 
- 2(\psi_+ \bar \psi_- + \psi_- \bar \psi_+ +\chi_+ \bar \chi_- + \chi_- \bar \chi_+)\L_+ \nonu \\
&+& 2(\psi_+ \bar \chi_+ + \psi_- \bar \chi_- +\chi_+ \bar \psi_+ + \chi_- \bar \psi_-)\D_- 
+ 2(\psi_+ \bar \chi_- + \psi_- \bar \chi_+ +\chi_+ \bar \psi_- + \chi_- \bar \psi_+)\D_+ 
\nonu \\
\label{b1}
\er
where
\br
\chi_{\pm}= \chi_1 \pm \chi_2, \quad \quad \bar \chi_{\pm} = \bar \chi_1 \pm \bar \chi_2, \quad \quad \varphi_{\pm} = \varphi_1 \pm  \varphi_2,  \cdots {\rm etc}
\nonu
\er 
and
\br
\L_{\pm} &=& \cosh ({{\phi_+ + \phi_-}\o 2}) \cosh ({{\varphi_+ + \varphi_-}\o 2}) \pm \cosh ({{\phi_+ - \phi_-}\o 2}) \cosh ({{\varphi_+ - \varphi_-}\o 2}), \nonu \\
\D_{\pm} &=& \sinh ({{\phi_+ + \phi_-}\o 2}) \sinh ({{\varphi_+ + \varphi_-}\o 2}) \pm \sinh ({{\phi_+ - \phi_-}\o 2}) \sinh ({{\varphi_+ - \varphi_-}\o 2}), \nonu \\ \nonu
\er
The solution of (\ref{b0}) and (\ref{b1}) is 
\br
B_0^{(+)}  &=& B_0^{(+)} (\phi_+, \varphi_+) = {{2 \a_3}\o {\a_2}}\( \cosh \phi_+ -  \cosh \varphi_+\), \label{b01}\\
B_0^{(-)}  &=& B_0^{(-)} (\phi_-, \varphi_-) = {{2 \a_2}\o {\a_3}} \(\cosh \phi_- -  \cosh \varphi_-\), \label{b02}
\er
and 
\br
B_1^{(+)}  &=& B_1^{(+)} (\phi_+, \varphi_+, \psi_+, \chi_+, f_1,f_2) \nonu \\
&=& {{i}\o {\sqrt{2}}}f_1\(-\a_3 (\psi_+ - \chi_+)\cosh {1\o 2} (\phi_+ + \varphi_+) \) \nonu \\
  &+&  {{i}\o {\sqrt{2}}}f_2\( \frac{8}{\alpha_2} (\psi_+ + \chi_+)\cosh {1\o 2} (\phi_+ - \varphi_+)\), \label{b11}
 \er
 \br
B_1^{(-)}  &=& B_1^{(-)} (\phi_-, \varphi_-, \bar \psi_-, \bar \chi_-, f_1,f_2) \nonu \\ 
&=&  {{i}\o {\sqrt{2}}}f_1\(\a_2 (\bar \psi_- + \bar \chi_-)\cosh {1\o 2} (\phi_- - \varphi_-) \) \nonu \\
  &+&  {{i}\o {\sqrt{2}}}f_2\( -\frac{8}{\alpha_3} (\bar \psi_- - \bar \chi_-)\cosh {1\o 2} (\phi_- + \varphi_-) \) 
   \label{b12}
\er
where $\a_2$ and $\a_3$ are arbitrary constants.

These  results are consistent with the Backlund transformation  for the $N=2$ super sinh-Gordon obtained from a superfield formalism \cite{inprep}.  We should mention that the Backlund transformation (\ref{d}) with $B_0$ and $B_1$ given by (\ref{b01})-(\ref{b12}) are invariant under the supersymmetry transformation (\ref{s})-(\ref{s1}). In \cite{inprep}  some examples of Backlund solutions are discussed also. 

\vskip 1cm

 \noindent
{\bf Acknowledgements} \\
\vskip .1cm \noindent
{  We are grateful to  CNPq and FAPESP  for 
financial support.}

\end{document}